\documentstyle[12pt]{article}  
\begin{document}
\ifx\undefined\psfig\else\endinput\fi

%
\edef\psfigRestoreAt{\catcode`@=\number\catcode`@\relax}
\catcode`\@=11\relax
\newwrite\@unused
\def\ps@typeout#1{{\let\protect\string\immediate\write\@unused{#1}}}
\ps@typeout{psfig/tex 1.8}


\def\figurepath{./}
\def\psfigurepath#1{\edef\figurepath{#1}}

%
%
\def\@nnil{\@nil}
\def\@empty{}
\def\@psdonoop#1\@@#2#3{}
\def\@psdo#1:=#2\do#3{\edef\@psdotmp{#2}\ifx\@psdotmp\@empty \else
    \expandafter\@psdoloop#2,\@nil,\@nil\@@#1{#3}\fi}
\def\@psdoloop#1,#2,#3\@@#4#5{\def#4{#1}\ifx #4\@nnil \else
       #5\def#4{#2}\ifx #4\@nnil \else#5\@ipsdoloop #3\@@#4{#5}\fi\fi}
\def\@ipsdoloop#1,#2\@@#3#4{\def#3{#1}\ifx #3\@nnil 
       \let\@nextwhile=\@psdonoop \else
      #4\relax\let\@nextwhile=\@ipsdoloop\fi\@nextwhile#2\@@#3{#4}}
\def\@tpsdo#1:=#2\do#3{\xdef\@psdotmp{#2}\ifx\@psdotmp\@empty \else
    \@tpsdoloop#2\@nil\@nil\@@#1{#3}\fi}
\def\@tpsdoloop#1#2\@@#3#4{\def#3{#1}\ifx #3\@nnil 
       \let\@nextwhile=\@psdonoop \else
      #4\relax\let\@nextwhile=\@tpsdoloop\fi\@nextwhile#2\@@#3{#4}}
%
\ifx\undefined\fbox
\newdimen\fboxrule
\newdimen\fboxsep
\newdimen\ps@tempdima
\newbox\ps@tempboxa
\fboxsep = 3pt
\fboxrule = .4pt
\long\def\fbox#1{\leavevmode\setbox\ps@tempboxa\hbox{#1}\ps@tempdima\fboxrule
    \advance\ps@tempdima \fboxsep \advance\ps@tempdima \dp\ps@tempboxa
   \hbox{\lower \ps@tempdima\hbox
  {\vbox{\hrule height \fboxrule
          \hbox{\vrule width \fboxrule \hskip\fboxsep
          \vbox{\vskip\fboxsep \box\ps@tempboxa\vskip\fboxsep}\hskip 
                 \fboxsep\vrule width \fboxrule}
                 \hrule height \fboxrule}}}}
\fi
%
%
\newread\ps@stream
\newif\ifnot@eof       
\newif\if@noisy        
\newif\if@atend        
\newif\if@psfile       
%
%
{\catcode`\%=12\global\gdef\epsf@start{
\def\epsf@PS{PS}
\def\epsf@getbb#1{%
%
%
\openin\ps@stream=#1
\ifeof\ps@stream\ps@typeout{Error, File #1 not found}\else
%
%
   {\not@eoftrue \chardef\other=12
    \def\do##1{\catcode`##1=\other}\dospecials \catcode`\ =10
    \loop
       \if@psfile
	  \read\ps@stream to \epsf@fileline
       \else{
	  \obeyspaces
          \read\ps@stream to \epsf@tmp\global\let\epsf@fileline\epsf@tmp}
       \fi
       \ifeof\ps@stream\not@eoffalse\else
%
%
       \if@psfile\else
       \expandafter\epsf@test\epsf@fileline:. \\%
       \fi
%
%
          \expandafter\epsf@aux\epsf@fileline:. \\%
       \fi
   \ifnot@eof\repeat
   }\closein\ps@stream\fi}%
%
%
\long\def\epsf@test#1#2#3:#4\\{\def\epsf@testit{#1#2}
			\ifx\epsf@testit\epsf@start\else
\ps@typeout{Warning! File does not start with `\epsf@start'.  It may not be a PostScript file.}
			\fi
			\@psfiletrue} 
%
%
{\catcode`\%=12\global\let\epsf@percent=
%
%
%
\long\def\epsf@aux#1#2:#3\\{\ifx#1\epsf@percent
   \def\epsf@testit{#2}\ifx\epsf@testit\epsf@bblit
	\@atendfalse
        \epsf@atend #3 . \\%
	\if@atend	
	   \if@verbose{
		\ps@typeout{psfig: found `(atend)'; continuing search}
	   }\fi
        \else
        \epsf@grab #3 . . . \\%
        \not@eoffalse
        \global\no@bbfalse
        \fi
   \fi\fi}%
%
%
\def\epsf@grab #1 #2 #3 #4 #5\\{%
   \global\def\epsf@llx{#1}\ifx\epsf@llx\empty
      \epsf@grab #2 #3 #4 #5 .\\\else
   \global\def\epsf@lly{#2}%
   \global\def\epsf@urx{#3}\global\def\epsf@ury{#4}\fi}%
%
%
\def\epsf@atendlit{(atend)} 
\def\epsf@atend #1 #2 #3\\{%
   \def\epsf@tmp{#1}\ifx\epsf@tmp\empty
      \epsf@atend #2 #3 .\\\else
   \ifx\epsf@tmp\epsf@atendlit\@atendtrue\fi\fi}


\chardef\letter = 11
\chardef\other = 12

\newif \ifdebug 
\newif\ifc@mpute 
\c@mputetrue 

\let\then = \relax
\def\r@dian{pt }
\let\r@dians = \r@dian
\let\dimensionless@nit = \r@dian
\let\dimensionless@nits = \dimensionless@nit
\def\internal@nit{sp }
\let\internal@nits = \internal@nit
\newif\ifstillc@nverging
\def \Mess@ge #1{\ifdebug \then \message {#1} \fi}

{ 
	\catcode `\@ = \letter
	\gdef \nodimen {\expandafter \n@dimen \the \dimen}
	\gdef \term #1 #2 #3%
	       {\edef \t@ {\the #1}
		\edef \t@@ {\expandafter \n@dimen \the #2\r@dian}%
		\t@rm {\t@} {\t@@} {#3}%
	       }
	\gdef \t@rm #1 #2 #3%
	       {{%
		\count 0 = 0
		\dimen 0 = 1 \dimensionless@nit
		\dimen 2 = #2\relax
		\Mess@ge {Calculating term #1 of \nodimen 2}%
		\loop
		\ifnum	\count 0 < #1
		\then	\advance \count 0 by 1
			\Mess@ge {Iteration \the \count 0 \space}%
			\Multiply \dimen 0 by {\dimen 2}%
			\Mess@ge {After multiplication, term = \nodimen 0}%
			\Divide \dimen 0 by {\count 0}%
			\Mess@ge {After division, term = \nodimen 0}%
		\repeat
		\Mess@ge {Final value for term #1 of 
				\nodimen 2 \space is \nodimen 0}%
		\xdef \Term {#3 = \nodimen 0 \r@dians}%
		\aftergroup \Term
	       }}
	\catcode `\p = \other
	\catcode `\t = \other
	\gdef \n@dimen #1pt{#1} 
}

\def \Divide #1by #2{\divide #1 by #2} 

\def \Multiply #1by #2
       {{
	\count 0 = #1\relax
	\count 2 = #2\relax
	\count 4 = 65536
	\Mess@ge {Before scaling, count 0 = \the \count 0 \space and
			count 2 = \the \count 2}%
	\ifnum	\count 0 > 32767 
	\then	\divide \count 0 by 4
		\divide \count 4 by 4
	\else	\ifnum	\count 0 < -32767
		\then	\divide \count 0 by 4
			\divide \count 4 by 4
		\else
		\fi
	\fi
	\ifnum	\count 2 > 32767 
	\then	\divide \count 2 by 4
		\divide \count 4 by 4
	\else	\ifnum	\count 2 < -32767
		\then	\divide \count 2 by 4
			\divide \count 4 by 4
		\else
		\fi
	\fi
	\multiply \count 0 by \count 2
	\divide \count 0 by \count 4
	\xdef \product {#1 = \the \count 0 \internal@nits}%
	\aftergroup \product
       }}

\def\r@duce{\ifdim\dimen0 > 90\r@dian \then   
		\multiply\dimen0 by -1
		\advance\dimen0 by 180\r@dian
		\r@duce
	    \else \ifdim\dimen0 < -90\r@dian \then  
		\advance\dimen0 by 360\r@dian
		\r@duce
		\fi
	    \fi}

\def\Sine#1%
       {{%
	\dimen 0 = #1 \r@dian
	\r@duce
	\ifdim\dimen0 = -90\r@dian \then
	   \dimen4 = -1\r@dian
	   \c@mputefalse
	\fi
	\ifdim\dimen0 = 90\r@dian \then
	   \dimen4 = 1\r@dian
	   \c@mputefalse
	\fi
	\ifdim\dimen0 = 0\r@dian \then
	   \dimen4 = 0\r@dian
	   \c@mputefalse
	\fi
	\ifc@mpute \then
		\divide\dimen0 by 180
		\dimen0=3.141592654\dimen0
		\dimen 2 = 3.1415926535897963\r@dian 
		\divide\dimen 2 by 2 
		\Mess@ge {Sin: calculating Sin of \nodimen 0}%
		\count 0 = 1 
		\dimen 2 = 1 \r@dian 
		\dimen 4 = 0 \r@dian 
		\loop
			\ifnum	\dimen 2 = 0 
			\then	\stillc@nvergingfalse 
			\else	\stillc@nvergingtrue
			\fi
			\ifstillc@nverging 
			\then	\term {\count 0} {\dimen 0} {\dimen 2}%
				\advance \count 0 by 2
				\count 2 = \count 0
				\divide \count 2 by 2
				\ifodd	\count 2 
				\then	\advance \dimen 4 by \dimen 2
				\else	\advance \dimen 4 by -\dimen 2
				\fi
		\repeat
	\fi		
			\xdef \sine {\nodimen 4}%
       }}

\def\Cosine#1{\ifx\sine\UnDefined\edef\Savesine{\relax}\else
		             \edef\Savesine{\sine}\fi
	{\dimen0=#1\r@dian\advance\dimen0 by 90\r@dian
	 \Sine{\nodimen 0}
	 \xdef\cosine{\sine}
	 \xdef\sine{\Savesine}}}	      

\def\psdraft{
	\def\@psdraft{0}
}
\def\psfull{
	\def\@psdraft{100}
}

\psfull

\newif\if@scalefirst
\def\psscalefirst{\@scalefirsttrue}
\def\psrotatefirst{\@scalefirstfalse}
\psrotatefirst

\newif\if@draftbox
\def\psnodraftbox{
	\@draftboxfalse
}
\def\psdraftbox{
	\@draftboxtrue
}
\@draftboxtrue

\newif\if@prologfile
\newif\if@postlogfile
\def\pssilent{
	\@noisyfalse
}
\def\psnoisy{
	\@noisytrue
}
\psnoisy
\newif\if@bbllx
\newif\if@bblly
\newif\if@bburx
\newif\if@bbury
\newif\if@height
\newif\if@width
\newif\if@rheight
\newif\if@rwidth
\newif\if@angle
\newif\if@clip
\newif\if@verbose
\def\@p@@sclip#1{\@cliptrue}

\newif\if@decmpr


\def\@p@@sfigure#1{\def\@p@sfile{null}\def\@p@sbbfile{null}
	        \openin1=#1.bb
		\ifeof1\closein1
	        	\openin1=\figurepath#1.bb
			\ifeof1\closein1
			        \openin1=#1
				\ifeof1\closein1%
				       \openin1=\figurepath#1
					\ifeof1
					   \ps@typeout{Error, File #1 not found}
						\if@bbllx\if@bblly
				   		\if@bburx\if@bbury
			      				\def\@p@sfile{#1}%
			      				\def\@p@sbbfile{#1}%
							\@decmprfalse
				  	   	\fi\fi\fi\fi
					\else\closein1
				    		\def\@p@sfile{\figurepath#1}%
				    		\def\@p@sbbfile{\figurepath#1}%
						\@decmprfalse
	                       		\fi%
			 	\else\closein1%
					\def\@p@sfile{#1}
					\def\@p@sbbfile{#1}
					\@decmprfalse
			 	\fi
			\else
				\def\@p@sfile{\figurepath#1}
				\def\@p@sbbfile{\figurepath#1.bb}
				\@decmprtrue
			\fi
		\else
			\def\@p@sfile{#1}
			\def\@p@sbbfile{#1.bb}
			\@decmprtrue
		\fi}

\def\@p@@sfile#1{\@p@@sfigure{#1}}

\def\@p@@sbbllx#1{
		\@bbllxtrue
		\dimen100=#1
		\edef\@p@sbbllx{\number\dimen100}
}
\def\@p@@sbblly#1{
		\@bbllytrue
		\dimen100=#1
		\edef\@p@sbblly{\number\dimen100}
}
\def\@p@@sbburx#1{
		\@bburxtrue
		\dimen100=#1
		\edef\@p@sbburx{\number\dimen100}
}
\def\@p@@sbbury#1{
		\@bburytrue
		\dimen100=#1
		\edef\@p@sbbury{\number\dimen100}
}
\def\@p@@sheight#1{
		\@heighttrue
		\dimen100=#1
   		\edef\@p@sheight{\number\dimen100}
}
\def\@p@@swidth#1{
		\@widthtrue
		\dimen100=#1
		\edef\@p@swidth{\number\dimen100}
}
\def\@p@@srheight#1{
		\@rheighttrue
		\dimen100=#1
		\edef\@p@srheight{\number\dimen100}
}
\def\@p@@srwidth#1{
		\@rwidthtrue
		\dimen100=#1
		\edef\@p@srwidth{\number\dimen100}
}
\def\@p@@sangle#1{
		\@angletrue
		\edef\@p@sangle{#1} 
}
\def\@p@@ssilent#1{ 
		\@verbosefalse
}
\def\@p@@sprolog#1{\@prologfiletrue\def\@prologfileval{#1}}
\def\@p@@spostlog#1{\@postlogfiletrue\def\@postlogfileval{#1}}
\def\@cs@name#1{\csname #1\endcsname}
\def\@setparms#1=#2,{\@cs@name{@p@@s#1}{#2}}
%
%
\def\ps@init@parms{
		\@bbllxfalse \@bbllyfalse
		\@bburxfalse \@bburyfalse
		\@heightfalse \@widthfalse
		\@rheightfalse \@rwidthfalse
		\def\@p@sbbllx{}\def\@p@sbblly{}
		\def\@p@sbburx{}\def\@p@sbbury{}
		\def\@p@sheight{}\def\@p@swidth{}
		\def\@p@srheight{}\def\@p@srwidth{}
		\def\@p@sangle{0}
		\def\@p@sfile{} \def\@p@sbbfile{}
		\def\@p@scost{10}
		\def\@sc{}
		\@prologfilefalse
		\@postlogfilefalse
		\@clipfalse
		\if@noisy
			\@verbosetrue
		\else
			\@verbosefalse
		\fi
}
%
%
\def\parse@ps@parms#1{
	 	\@psdo\@psfiga:=#1\do
		   {\expandafter\@setparms\@psfiga,}}
%
%
\newif\ifno@bb
\def\bb@missing{
	\if@verbose{
		\ps@typeout{psfig: searching \@p@sbbfile \space  for bounding box}
	}\fi
	\no@bbtrue
	\epsf@getbb{\@p@sbbfile}
        \ifno@bb \else \bb@cull\epsf@llx\epsf@lly\epsf@urx\epsf@ury\fi
}	
\def\bb@cull#1#2#3#4{
	\dimen100=#1 bp\edef\@p@sbbllx{\number\dimen100}
	\dimen100=#2 bp\edef\@p@sbblly{\number\dimen100}
	\dimen100=#3 bp\edef\@p@sbburx{\number\dimen100}
	\dimen100=#4 bp\edef\@p@sbbury{\number\dimen100}
	\no@bbfalse
}
\newdimen\p@intvaluex
\newdimen\p@intvaluey
\def\rotate@#1#2{{\dimen0=#1 sp\dimen1=#2 sp
		  \global\p@intvaluex=\cosine\dimen0
		  \dimen3=\sine\dimen1
		  \global\advance\p@intvaluex by -\dimen3
		  \global\p@intvaluey=\sine\dimen0
		  \dimen3=\cosine\dimen1
		  \global\advance\p@intvaluey by \dimen3
		  }}
\def\compute@bb{
		\no@bbfalse
		\if@bbllx \else \no@bbtrue \fi
		\if@bblly \else \no@bbtrue \fi
		\if@bburx \else \no@bbtrue \fi
		\if@bbury \else \no@bbtrue \fi
		\ifno@bb \bb@missing \fi
		\ifno@bb \ps@typeout{FATAL ERROR: no bb supplied or found}
			\no-bb-error
		\fi
		%
%
		\count203=\@p@sbburx
		\count204=\@p@sbbury
		\advance\count203 by -\@p@sbbllx
		\advance\count204 by -\@p@sbblly
		\edef\ps@bbw{\number\count203}
		\edef\ps@bbh{\number\count204}
		\if@angle 
			\Sine{\@p@sangle}\Cosine{\@p@sangle}
	        	{\dimen100=\maxdimen\xdef\r@p@sbbllx{\number\dimen100}
					    \xdef\r@p@sbblly{\number\dimen100}
			                    \xdef\r@p@sbburx{-\number\dimen100}
					    \xdef\r@p@sbbury{-\number\dimen100}}
%
                        \def\minmaxtest{
			   \ifnum\number\p@intvaluex<\r@p@sbbllx
			      \xdef\r@p@sbbllx{\number\p@intvaluex}\fi
			   \ifnum\number\p@intvaluex>\r@p@sbburx
			      \xdef\r@p@sbburx{\number\p@intvaluex}\fi
			   \ifnum\number\p@intvaluey<\r@p@sbblly
			      \xdef\r@p@sbblly{\number\p@intvaluey}\fi
			   \ifnum\number\p@intvaluey>\r@p@sbbury
			      \xdef\r@p@sbbury{\number\p@intvaluey}\fi
			   }
			\rotate@{\@p@sbbllx}{\@p@sbblly}
			\minmaxtest
			\rotate@{\@p@sbbllx}{\@p@sbbury}
			\minmaxtest
			\rotate@{\@p@sbburx}{\@p@sbblly}
			\minmaxtest
			\rotate@{\@p@sbburx}{\@p@sbbury}
			\minmaxtest
			\edef\@p@sbbllx{\r@p@sbbllx}\edef\@p@sbblly{\r@p@sbblly}
			\edef\@p@sbburx{\r@p@sbburx}\edef\@p@sbbury{\r@p@sbbury}
		\fi
		\count203=\@p@sbburx
		\count204=\@p@sbbury
		\advance\count203 by -\@p@sbbllx
		\advance\count204 by -\@p@sbblly
		\edef\@bbw{\number\count203}
		\edef\@bbh{\number\count204}
}
%
%
\def\in@hundreds#1#2#3{\count240=#2 \count241=#3
		     \count100=\count240	
		     \divide\count100 by \count241
		     \count101=\count100
		     \multiply\count101 by \count241
		     \advance\count240 by -\count101
		     \multiply\count240 by 10
		     \count101=\count240	
		     \divide\count101 by \count241
		     \count102=\count101
		     \multiply\count102 by \count241
		     \advance\count240 by -\count102
		     \multiply\count240 by 10
		     \count102=\count240	
		     \divide\count102 by \count241
		     \count200=#1\count205=0
		     \count201=\count200
			\multiply\count201 by \count100
		 	\advance\count205 by \count201
		     \count201=\count200
			\divide\count201 by 10
			\multiply\count201 by \count101
			\advance\count205 by \count201
		     \count201=\count200
			\divide\count201 by 100
			\multiply\count201 by \count102
			\advance\count205 by \count201
		     \edef\@result{\number\count205}
}
\def\compute@wfromh{
		\in@hundreds{\@p@sheight}{\@bbw}{\@bbh}
		\edef\@p@swidth{\@result}
}
\def\compute@hfromw{
	        \in@hundreds{\@p@swidth}{\@bbh}{\@bbw}
		\edef\@p@sheight{\@result}
}
\def\compute@handw{
		\if@height 
			\if@width
			\else
				\compute@wfromh
			\fi
		\else 
			\if@width
				\compute@hfromw
			\else
				\edef\@p@sheight{\@bbh}
				\edef\@p@swidth{\@bbw}
			\fi
		\fi
}
\def\compute@resv{
		\if@rheight \else \edef\@p@srheight{\@p@sheight} \fi
		\if@rwidth \else \edef\@p@srwidth{\@p@swidth} \fi
}
%
\def\compute@sizes{
	\compute@bb
	\if@scalefirst\if@angle
	\if@width
	   \in@hundreds{\@p@swidth}{\@bbw}{\ps@bbw}
	   \edef\@p@swidth{\@result}
	\fi
	\if@height
	   \in@hundreds{\@p@sheight}{\@bbh}{\ps@bbh}
	   \edef\@p@sheight{\@result}
	\fi
	\fi\fi
	\compute@handw
	\compute@resv}

%
%
\def\psfig#1{\vbox {
	%
	\ps@init@parms
	\parse@ps@parms{#1}
	\compute@sizes
	\ifnum\@p@scost<\@psdraft{
		\special{ps::[begin] 	\@p@swidth \space \@p@sheight \space
				\@p@sbbllx \space \@p@sbblly \space
				\@p@sbburx \space \@p@sbbury \space
				startTexFig \space }
		\if@angle
			\special {ps:: \@p@sangle \space rotate \space} 
		\fi
		\if@clip{
			\if@verbose{
				\ps@typeout{(clip)}
			}\fi
			\special{ps:: doclip \space }
		}\fi
		\if@prologfile
		    \special{ps: plotfile \@prologfileval \space } \fi
		\if@decmpr{
			\if@verbose{
				\ps@typeout{psfig: including \@p@sfile.Z \space }
			}\fi
			\special{ps: plotfile "`zcat \@p@sfile.Z" \space }
		}\else{
			\if@verbose{
				\ps@typeout{psfig: including \@p@sfile \space }
			}\fi
			\special{ps: plotfile \@p@sfile \space }
		}\fi
		\if@postlogfile
		    \special{ps: plotfile \@postlogfileval \space } \fi
		\special{ps::[end] endTexFig \space }
		\vbox to \@p@srheight true sp{
			\hbox to \@p@srwidth true sp{
				\hss
			}
		\vss
		}
	}\else{
		\if@draftbox{		
			\hbox{\frame{\vbox to \@p@srheight true sp{
			\vss
			\hbox to \@p@srwidth true sp{ \hss \@p@sfile \hss }
			\vss
			}}}
		}\else{
			\vbox to \@p@srheight true sp{
			\vss
			\hbox to \@p@srwidth true sp{\hss}
			\vss
			}
		}\fi

	}\fi
}}
\psfigRestoreAt

\pagestyle{plain}\textheight=21truecm\textwidth=13truecm
\begin{center}
\begin{bf}
ON THE NON-LINEAR 
HYDRODYNAMIC STABILITY OF THIN KEPLERIAN DISKS
\end{bf}  
\end{center}
\vspace{1.cm}  
\begin{center}
Patrick Godon and Mario Livio \\
\vspace{5.cm}  
Space Telescope Science Institute \\
3700 San Martin Drive \\ 
Baltimore, MD 21218 \\
\vspace{0.5cm}  
e-mail: godon@stsci.edu, mlivio@stsci.edu  \\ 
\end{center}
\newpage
\baselineskip 15pt plus 2pt
\begin{center}
\bf{Abstract}
\end{center}
The non-linear hydrodynamic stability of thin, compressible,
Keplerian disks is studied on the large two-dimensional
compressible scale,    
using a high-order accuracy spectral method.  
We find that purely hydrodynamical perturbations, can 
develop initially into either sheared disturbances or coherent
vortices. However, the perturbations decay and do not
evolve into a self-sustained turbulence. 
Temporarily, due to an inverse cascade of energy, which is 
characteristic of two-dimensional flows, energy is being transferred
to the largest scale mode before being dissipated. In the case
of an inner reflecting boundary condition, it is found that
the innermost disk is globally unstable to non-axisymmetric
modes which can evolve into turbulence. However, this turbulence 
cannot play a significant role in angular momentum transport in disks. 
 \\ \\  

Subject Headings: accretion, accretion disks -
hydrodynamics - instabilities - turbulence   

\newpage 
\section{Introduction}

Recently, it has been demonstrated (Balbus \& Hawley 1991;  
see also their review Balbus \& Hawley 1998) 
that the origin of turbulence and angular 
momentum transport in accretion disks is very likely related to 
a linear, local, magnetohydrodynamic (MHD)
instability-the magnetorotational 
instability (Velikhov 1959; Chandrasekhar 1960; Balbus \& Hawley 1991).
This instability has been shown to successfully reproduce the 
order of magnitude of the effective viscosity in ionized
disks (with a viscosity parameter $\alpha \approx 0.001-0.1$; 
Hawley, Gammie \& Balbus 1995, Brandenburg et al. 1995; 
Vishniac \& Brandenburg 1997). 
The situation in dense and cool disks, like the ones found in 
Young Stellar Objects, is somewhat more ambiguous (e.g.
Blaes \& Balbus 1994; Reg\"os 1997) and it has been suggested
that the source of turbulence and angular momentum in these
systems might be due to a hydrodynamical instability (e.g.
Zahn 1990; Dubrulle 1993). 
However, no linear hydrodynamic 
instability has been identified (e.g. Stewart 1975), and  
efforts have  
focused on nonlinear instabilities (e.g. Zahn 1990; Dubrulle \& Knobloch 1992; 
Dubrulle 1992, 1993). The latter authors have derived 
some interesting properties of the turbulence (and the viscosity in the solar 
nebula), however, they actually {\it {assumed}} the turbulence to exist 
{\it{ab initio}}, rather than rigorously demonstrating that it is
the result of some instability (basing their assumption on the fact
that Couette and Poisseuille flows do exhibit 
nonlinear instabilities in
laboratory experiments; e.g. Coles 1965; Daviaud, 
Hegseth \& Berg\'e 1992; see also Zahn 1990).
In fact, until now, it has not been shown that a Keplerian disk is 
unstable to a local nonlinear hydrodynamic instability that leads to
turbulence. Quite to the contrary, using a relatively 
low-resolution finite-difference code, Balbus, Hawley \& Stone (1996) 
have shown that Keplerian flows are stable to finite amplitude 
perturbations on the small three-dimensional scale,
using the shearing-box approximation.
They explain their numerical findings by pointing out that  
the epicyclic
term is always a sink in the energy equation, and consequently, the 
Reynolds stress is expected to decay, even if turbulence transports
angular momentum outwards. However, no rigorous analytical proof
is provided. \\ 

In order to further test the hydrodynamic
stability of Keplerian disks, we propose in this work to check the 
non-linear stability of the flow on the {\it large} compressible 
two-dimensional scale, since on the small three-dimensional scale
the flow is expected to be stable (Balbus et al. 1996). 
The motivations behind this particular calculation are multiple: 
(i) two-dimensional compressible rotating flows (e.g. the shallow
water approximation) do exhibit a transition to turbulence due to the
propagation of waves (e.g. Pedlosky 1987; Cushman-Roisin 1994;
see also Cho \& Polvani 1996a,b, and see \S 2.5).
(ii) The wavelength and the amplitude of the perturbation needed 
for a subcritical transition to turbulence 
are not known a priori. (iii)
The present method of calculation allows us to study the {\it entire 
disk}, thus including the effects of the amplification 
of propagating waves,  
rotation, curvature and compressibility, which might have  
crucial effects on the flow (see e.g. Cho \& Polvani 1996a
for such a discussion). \\ 

Since the transition to turbulence in shear 
flows is usually assumed to be a strictly three-dimensional
effect (e.g. Balbus \& Hawley 1998),  
we discuss in the next section the transition to turbulence
in three- as well as two-dimensional rotating flows. 
The equations and the numerical method are presented
in \S 3, the results are shown in \S 4, and a discussion follows.
    \\    

\section{Transition to turbulence}  

Planar shear flows are 
unstable to finite-amplitude perturbations only in three 
dimensions (3D) (e.g. Bayly, Orszag \& Herbert 1988),
and numerical simulations have also shown that a
subcritical transition to turbulence in the plane Couette 
flow occurs only
in 3D (e.g. Orszag \& Kells 1980; Dubrulle 1991). 
These results have led to
the general assumption that a hydrodynamic (subcritical) 
transition to turbulence 
in accretion disks can take place only in 3D.
However, this assumption is based entirely on experiments 
and simulations with {\it{incompressible}} 3D flows, and 
therefore great caution should be exercised in 
attempts to draw conclusions 
for 2D compressible flows in accretion disks.   
In particular, waves play a crucial role 
in the formation of turbulence in the  
shallow water approximation of planetary 
atmospheres (e.g. Cho \& Polvani 1996a, 1996b).
Therefore, there are good reasons to believe that
turbulence can in principle at least 
develop in two-dimensional disks as well. 
 \\  

In this section we review the subcritical transition to 
turbulence in three-dimensional incompressible 
and two-dimensional compressible flows. \\ \\ 
 
\subsection{Three-dimensional incompressible flows} 

The transition to turbulence in 
three-dimensional incompressible flows has  
often been considered to be due to an inflectional instability.  
However, more recently it has been realized  
(e.g. Yang 1987; Hamilton \& Abernathy 1994; 
Dauchot \& Daviaud 1995; Hegseth 1996) that 
the destabilization process that leads to turbulence
is associated with the presence of streamwise vortices
rather than with inflection points: non-transitional
flows were found to exhibit inflectional profiles as well.  
A streamwise vortex always induces inflections
in the streamwise velocity profiles, but the transition
to turbulence occurs only when the strength (circulation)
of the vortex is above some critical value 
(Hamilton \& Abernathy 1994). 
The turbulence is fed by the streamwise vortex
which transfers energy from the bulk motion of the flow 
to the turbulent region (usually a 'spot'). 
Streamwise vortices in the flow are purely three dimensional,  
and cannot be reproduced by two dimensional simulations.

\subsection{Two-dimensional compressible flows}  

Two-dimensional flows with inflectional velocity profiles   
satisfy the necessary (but not sufficient)  condition for 
instability (e.g. Fj{\o}rtoft 1950). 
Therefore, one of the features 
we might expect to observe if the flow is unstable, is the presence
of inflection points, since this is a necessary (but not sufficient)
condition for instability in two-dimensional flows.   

However, very little is known about the dynamics of
rotating flows when compressibility is important,
since only a few studies  
have been carried out for supersonic rotating flows 
in two dimensions 
(e.g. Farge \& Sadourny 1989; Tomasini, Dolez \& L\'eorat 1996;
Godon 1998). 
In their work on two-dimensional transonic shear flows, Tomasini
et al. (1996) did not rule out turbulence, although it was not
obtained numerically. The turbulence found in Godon (1998) was
the non-linear consequence of the Papaloizou-Pringle (1984)
instability, which itself was due to an inner reflecting boundary
condition (Gat \& Livio 1992).  

In the shallow water approximation of two-dimensional planetary
atmospheres, the shear flow is unstable
to wavelengths that are longer than a critical value. 
These waves have a phase
speed that matches the mean current velocity within the flow, which
permits a transfer of energy from the current to the
wave. This process is known in geophysical fluid dynamics as
{\it the barotropic instability} (e.g. Pedlosky 1987; Cushman-Roisin 1994).
This instability is linear, and it is responsible
for the formation of turbulence in the atmosphere on the
large two-dimensional scales which transport angular momentum. 
An important point to mention here is that  
two-dimensional compressible polytropic flows (with a polytropic index
n=1) 
can be represented by the shallow water equations, 
when the depth $h$ of the shallow water flow
is replaced by the surface density $\Sigma$,
and the surface gravity waves speed $\sqrt{gh}$ corresponds
to the sound speed (see e.g. Tomasini et al. 1996; Ingersoll \& Cho 1997, 
private communication; Bracco et al. 1998, 1999).   
Moreover, 
in the thin disk approximation, like
in the atmosphere of a planet, the long wavelength modes
are two-dimensional
because of the finite vertical extent of the flow and because
of the rotation. On the smaller scales (say $\lambda < H$ in disks
and $\lambda < h$ in the atmosphere) the subsonic 
flow is not affected by the rotation or the curvature 
(see also Dubrulle \& Valdettaro 1992). However, while there
are many similarities between the Shallow Water Equations and the
two-dimensional compressible equations for a disk, there are 
also some important differences (see \S 5). \\  

\section{Numerical modeling} 

\subsection{The Equations} 
The equations (e.g. Tassoul 1978) are written in a cylindrical
coordinates system $(r, \phi, z)$, and are further
integrated in the ($z$) vertical direction (e.g. Pringle 1981). 
We define a mean density ${\bar{\rho}}$ using the definition of 
the surface density  $\Sigma = \int_{-H}^{+H} \rho dz 
= \int_{-H_{max}}^{+H_{max}} {\bar{\rho}} dz$, where $H_{max}$ has  
a constant value $H_{max} > max(H)$. This relation always  
holds since by definition the integral vanishes
for $H > H_{max}$. The surface density $\Sigma$ is then substituted
in the equations by $2 H_{max} {\bar{\rho}} $ and the equations are further
divided by $2 H_{max}$. The equations are written for the density
$\rho$ (where for simplicity we have dropped the bar), 
the radial momentum $U = \rho v_r$ and the angular momentum
$W=\rho v_{\phi}$.
 \\ 
The equations are:  
the conservation of mass,  
\begin{equation}
\frac{\partial \rho}{\partial t} + 
\frac{1}{r} \frac{\partial }{\partial r}  r U +
\frac{1}{r} \frac{\partial }{\partial \phi} W =0,  
\end{equation}  
the conservation of momentum in the radial dimension,
\[ 
\frac{\partial U }{\partial t}  
+\frac{1}{r} \frac{\partial }{\partial r} r \left( 
  \frac{U^2}{\rho} + P - R_{rr} \right)   
+\frac{1}{r} \frac{\partial }{\partial \phi}  \left( 
  \frac{UW}{\rho} - R_{r \phi} \right)   \]  
\begin{equation}
= 
\frac{1}{r} \left( 
\frac{W^2}{\rho} -\rho \frac{GM}{r} +P -R_{\phi \phi} 
\right) , 
\end{equation}  
the conservation of momentum in the angular direction, 
\[ 
\frac{\partial W }{\partial t}  
+\frac{1}{r} \frac{\partial }{\partial r} r \left( 
  \frac{UW}{\rho} - R_{\phi r} \right)   
+\frac{1}{r} \frac{\partial }{\partial \phi}  \left( 
  \frac{W^2}{\rho} + P - R_{\phi \phi} \right)   \]  
\begin{equation}
=
\frac{1}{r} \left( 
-\frac{UW}{\rho} + R_{r \phi} 
\right) , 
\end{equation}  
where $R_{ij}$ ($i=r,\phi;j=r,\phi$) is the Reynolds
stress tensor.
\[ 
R_{rr} =2 \nu \rho \frac{\partial}{\partial r} 
\left( \frac{U}{\rho} \right),  
\] 
\[ 
R_{r \phi} = R_{\phi r}= \nu \rho \left[ 
r \frac{\partial}{\partial r} 
\left( \frac{U}{r \rho} \right) 
+ \frac{1}{r} \frac{\partial}{\partial \phi} 
\left( \frac{U}{\rho} \right) \right],  
\]
\[ 
R_{\phi \phi} = 2 \nu \rho \left[ \frac{U}{\rho r}
+ \frac{1}{r} \frac{\partial}{\partial \phi} \left(
\frac{W}{\rho} \right) \right].
\]
 
The pressure is given assuming a polytropic relation
\begin{equation}
P=K \rho^{1 +1/n}.
\end{equation}
We chose $n=3$ for 
the polytropic index, while the polytropic constant $K$ was fixed 
by choosing $H/r$ at the outer radial boundary. \\  

We use an alpha prescription (Shakura \& Sunyaev 1973)
for the viscosity law 
$( \nu = \alpha c_s H = \alpha c_s^2 / \Omega_K )$, 
where $c_s$ is the sound speed and $H = c_s/\Omega_K$ 
is the vertical height of the disk. 

The Reynolds number in the flow is given by 
\[ R = \frac{L v}{\nu}, \] 
where $L$ is a characteristic dimension of the computational domain,
$v$ is the velocity of the flow (or more precisely the velocity 
change in the flow over a distance $L$) 
and $\nu$ is the viscosity.
Since we are solving for the entire disk, $L \approx r$ and 
$v \approx v_K$, and the Reynolds number becomes
\[ R_{disk} = \alpha^{-1} (H/r)^{-2}. \]

\subsection{Numerical method}  

We use a Fourier-Chebyshev spectral method 
of collocation (described in Godon 1997)  
to study waves in accretion disks. 
Spectral Methods are global and of order 
$N/2$ in space (where $N$ is the number of
grid points; see also Gottlieb \& Orszag 1977; Voigt, 
Gottlieb \& Hussaini 1984; Canuto et al. 1988) 
and they make use of fast Fourier transforms.  
These methods are relatively fast and accurate 
and consequently they are frequently used to 
solve for turbulent flows (e.g. She, Jackson \& Orszag 1991;
Cho \& Polvani 1996a). 
All the details on the numerical method 
can be found in Godon (1997), here we only give 
a brief outline of the numerical modeling. We would like however
to remark that because spectral methods are high-order-accuracy methods,
the number of points needed to achieve
a given accuracy is much less than in usual finite difference
methods. For example, a modest number of 17 points in a spectral
code has the accuracy of a low order 
finite difference solver with
a 100 points (Gottlieb \& Orszag 1977; Voigt, 
Gottlieb \& Hussaini 1984; Canuto et al. 1988).   
It is quite remarkable that 
Orszag \& Kells (1980) were able to study the subcritical 
transition to turbulence in the Plane Couette Flow with
a spectral method  
with a resolution as low as $16^3$ points. 
Spectral methods reach a high accuracy for smooth solutions, and the series
do not converge for discontinuous solutions (e.g. very strong shocks
in the flow can be responsible for numerical oscillations, the Gibbs
phenomenon).
However, in the present work the shocks that may form are due to the 
amplification of waves propagating inwards in the disk. These shocks
are (relatively) weak and they 
do not affect the accuracy of the results
presented here. 
For example, similar shocks
were obtained due to the amplification of tidally induced spiral waves
propagating inwards in the disk (with a Mach number of 1.3, 
Godon 1997).
These waves were found to reach
the inner boundary, even assuming an alpha viscosity parameter
$\alpha = 0.01$ and a resolution of 128x32 (Godon 1997), while in 
{\it inviscid} models using finite difference schemes (e.g. Savonije, Papaloizou
\& Lin 1994) 
the tidal waves were damped {\it before} reaching the inner boundary
even though the resolution was higher (200x64).
These results show that in the circumstances that apply to the 
present calculations (perturbations of a disk)
our spectral code has a higher accuracy than usual finite difference
codes with the same number of grid points.   
\\   

A Spectral filter was used to cut-off 
high frequencies. This implementation was used for numerical
convenience. It eliminates the high frequencies (e.g. due to 
shocks and aliasing errors of the non-linear terms) which can cause
numerical instability, while keeping a high enough number of 
terms in the spectral expansion to resolve the fine structure of 
the flow (Godon 1997, 1998; see also Don \& Gottlieb 1990). \\  

In all the numerical simulations, the alpha viscosity prescription 
was introduced in order 
to dissipate the energy in the small scales (say,   
for wavenumbers $k > k_{\nu}$),  
and the spectral filter was used to cut-off the highest frequencies
(for wavenumbers $k > k_0 > k_{\nu}$) which can grow due
to aliasing errors and the Gibbs phenomenon (Gottlieb \& Orszag
1977; Voigt et al. 1984; Canuto et al. 1988). To make sure
that the flow is numerically stable we check the energy spectra
($E_k$) of the flow and make sure that the energy decreases
(say, like $E_k \propto k^{-\beta}$ with $ 3 \le \beta \le 4$) 
for wavenumbers $k > 3M/4$,
where $M$ is the total number of modes (i.e. $k=M$ corresponds 
to the 
highest frequency). For low amplitude perturbations, 
a spectral filter is sufficient to damp the high
frequencies (by choosing $k_0 = 3M/4$). However, when
the perturbations are {\it strong} (like in the present work),
a small viscosity is needed in addition to the spectral
filters (the exponential cut-off filter used 
here is a {\it{weak}} filter).  

\section{Results} 

We have run several classes of models. In models of the first
class, we simply perturbed the disk with a finite-amplitude
perturbation, and free boundary conditions were applied at 
the inner and outer edges. In the second class of models, in 
addition to perturbing the disk, we also included (throughout
the run) the tidal effects from a companion star. In the third
class we introduced in the equations an artificial change
in the epicyclic term (see \S 4.3), and finally,
in the fourth class, we applied a reflective
boundary condition at the inner edge of the disk. In all the models
the initial rotation law is Keplerian and the radial velocity is 
set to zero. The initial density profile was chosen to fit the 
standard thin disk model (Shakura \& Sunyaev, 1973; 
i.e. $\rho (r) \propto r^{-15/8}$), but 
models with a flat density profile were also run with similar results.

\subsection{Finite perturbations, free boundaries} 

In this class of models, we
have run more than 10 models, changing the resolution
($N \times Q$, where $N$ is the number of collocation points 
in the Chebyshev expansion for the radial dimension, 
and $Q$ is the number of points in the Fourier expansion 
for the azimuthal dimension),
the ratio of the outer to the inner radius
($R_{out}/R_{in}$), the viscosity parameter ($\alpha$)
and the Mach number in the flow (${\cal{M}}=(H/r)^{-1}$).
The viscosity
parameter was varied from $\alpha = 0.1$ down to
$\alpha = 5 \times 10^{-5}$ (more than three orders
of magnitude change in the Reynolds number) 
and we tried mainly two
values of the Mach number, corresponding to
$H/r=0.05$ and $H/r=0.15$.
Not all the computed models are presented here.   \\ 

The optimal results (in terms of accuracy and speed) 
were obtained for
$R_{out}/R_{in} = 2$ and $N \times Q=128 \times 128$. 
Given the fact that we are 
working with a high-order accuracy spectral method, the 
resolution achieved with $128 \times 128$ points is equivalent
to that obtained by a usual second order accuracy finite-difference
method with more than $500^2$ grid points (Canuto et al. 1988).
This improved resolution is achieved
because the spatial resolution is not only a direct 
function of the number of grid points, but it is also an inverse  
function of the numerical dissipation, which is responsible for 
the flattening and spreading of fine structures in the flow.
High order accuracy spectral methods have very little numerical
dissipation in comparison to standard low order accuracy finite-difference
methods with the same number of grid points.  \\  
 
We chose in the present work to perturb the disk
velocities with a high order mode.
The initial perturbation was randomly chosen 
to have a maximum amplitude 
of about $0.1-0.2$ of the sound speed. 
However, even larger amplitude
perturbations were obtained naturally, due to 
the amplification of the waves as they
propagated inward into the disk, resulting in some 
cases in shocks in the inner region of the
computational domain.  
The perturbations were placed on a grid as shown  
in figure 1. 
The kinetic energy of the initial perturbations 
ranged from $10^{-5}$ to $10^{-3}$ of   
the background kinetic energy of the Keplerian flow. \\  


{\small{\it Figure 1:
A grayscale of the density shows the location of the perturbations
induced in the density of the disk.   
}} \\  

In the first model (model 1) we chose $\alpha =0.1$ 
with $R_{out}/R_{in} = 5$, $H/r=0.05$
and a resolution of 64x64.
On a very short time scale the initial perturbation
was damped and, as expected for such a large viscosity, 
the disk reached a stable configuration even before a 
significant propagation of the waves has been noticed. \\  

In a second model (model 2) we reduced the viscosity
parameter to $\alpha = 0.01$ and changed the numerical
resolution to $N \times Q=128 \times 128$ with  
$R_{out}/R_{in} = 2$, keeping $H/r=0.05$. In this case
the perturbation had time to propagate and form spiral
waves sheared by the flow. However, the short wavelength
waves were rapidly damped (see figure 2). 
After about 5 Keplerian
orbits all the waves were smoothed out. We found that
in the models the waves dissipated on a time scale
$\tau_{dis} \approx (\tau_{\nu}P_K)^{1/2}$, where $\tau_{\nu}=r^2/\nu$
is the viscous time scale and $P_K=2 \pi /\Omega_K$ is the Keplerian
Period. \\

{\small{\it Figure 2: The initial perturbation of the disk
induces waves that form spirals due to the shear in the flow.
In this model $H/r = 1/20$ and the viscosity parameter is 
$\alpha=0.01$.
The waves are smoothed out and decay on a time
scale $\tau_{dis} \approx (\tau_{\nu} P_K)^{1/2}$, where 
$P_K=2 \pi / \Omega_K$ is the Keplerian period and 
$\tau_{\nu}=r^2/\nu$ is the viscous time scale. 
}}  \\

In model 3 we have further reduced the viscosity to
$\alpha = 10^{-3}$, while keeping all the other parameters 
constant. In this case the waves decayed on a much longer time
scale, and the shear had sufficient time to act strongly
on the waves. This resulted in the formation of 
tightly wound spirals in the 
disk (figure 3, the radial wavelengths are very short
and are barely visible in the inner disk).
Here again the waves dissipated after a characteristic 
time $\tau_{dis}$. \\  

 
{\small{\it Figure 3: The same as Fig.2, with  
$\alpha = 10^{-3}$ (see text).  
}}  \\


As we have further reduced the viscosity in the disk, the wavelengths
of the waves in the radial direction decreased to a point where a higher  
resolution was needed. At this stage we decided to increase the ratio
$H/r$ while keeping the same resolution. In model 4
we therefore took $H/r = 0.15$ and $\alpha = 5 \times 10^{-5}$. Here the 
largest waves
dissipated on a time scale of the order of dozens of orbits, and the
effect of the viscosity was to damp the shortest wavelengths first.
As the longer wavelength modes propagated inward, they amplified and formed 
shocks similar to tidally induced spiral waves.
The spiral waves were not tightly wound and the wavelength in the radial
direction was not very short due to the lower Mach number in the 
disk (Figs. 4 and 5). \\ 


{\small{\it Figure 4: The grayscale of the density is shown 
(at $t \approx 2$ orbits) for 
model 4 with a very low viscosity 
($\alpha = 5 \times 10^{-5}$) and a lower Mach number 
($H/r\approx 1/7$). 
}}  \\


{\small{\it Figure 5: A grayscale of the density for
model 4 around $t \approx 10$ orbits.  
The dissipation time of the perturbations is long, of the order of
dozens of orbits, however the shortest wavelength disturbances
disappear within about $10$ orbits (see text). 
}}  \\

As an additional check of the stability of the disk, 
we have also looked at the
absolute vorticity of the flow. It is well known from the theory of
2D turbulent flows that an inflection point in the vorticity is a 
necessary condition for the appearance of an instability.
The absolute vorticity of the flow is shown in figure 6, a few
orbits after the initial perturbation of the disk. The striking feature
in figure 6 is a multitude of Z-like shapes criss-crossing the whole disk. 
The Z-shape in the absolute vorticity is a characteristic signature of the
instability of sheared disturbances (e.g. Haynes 1987). However, while
this instability is usually driving turbulence in shear flows, in the 
present Keplerian case it eventually stabilizes, as shown in figure 7.
This strongly suggests that  
the instability is deriving its energy from the initial perturbation  
rather than from the Keplerian motion of the flow itself.  \\  


{\small{\it Figure 6: A grayscale of the absolute vorticity for
model 4 after a few orbits. The initial perturbations have evolved into
Z-shapes, characteristic of the instability of sheared disturbances.  
}}  \\

 
{\small{\it Figure 7: A grayscale of the absolute vorticity for
model 4 after about 10 orbits. The flow has stabilized completely. 
}}  \\

In all the preceding models, the size and amplitude of the perturbation
induced in the density were significant (Figure 5), especially due to
the amplification of waves propagating inward (forming shocks). 
However, the size and amplitude of the perturbation induced in the 
vorticity profile were indeed very small (Figure 6), and the dissipation
of the vorticy perturbation might be due to the shear and the viscosity
(Figure 11 shows that modes higher than $m \approx 20$ are quickly damped for 
$\alpha = 5 \times 10^{-5}$). 
It is important to note that 
the amplitude of the perturbation needed for the development 
of turbulence depends on the Reynolds number 
(e.g. Dauchot \& Daviaud 1994). 
In fact, the critical amplitude for the 
non-linear growth of disturbances in shear flows follows a 
scaling law which depends on the shape of the initial disturbance,
as a result of the competition between the non-linear growth 
of the perturbation 
and the viscous dissipation mechanism (e.g. Dubrulle \& Nazarenko 1994).
Therefore, for a given resolution, a given configuration may  
need a large amplitude
perturbation to give rise to an instability.  
Consequently, in order to check the effects of a large
perturbation in the vorticity field, we decided to run an additional model
in which the initial vorticity perturbation is large in both 
size and amplitude.   
The reasons behind this vorticity perturbation are multiple. First, 
two-dimensional flows are unstable to local extrema of the vorticity
(inflexion points) that can lead to turbulence. Second, two-dimensional
planetary atmospheric turbulent 
flows are usually perturbed and studied in the vorticy field rather
than in the velocity field. Lastly, recent numerical simulations
using the shallow water equations  
(Bracco et al. 1998, 1999) have shown that vortices can survive in a
Keplerian shear flow for many orbits. \\ 

In this model (model 5) we chose $H/r=0.15$, 
$\alpha = 1 \times 10^{-4}$ and a polytropic index $n=2.5$. 
The initial vorticity perturbation 
profile is shown in Figure 8.

 
{\small{\it Figure 8: A grayscale of the initial vorticity
perturbation. For clarity, the Keplerian velocity 
background has been subtracted from  
the angular velocity. The maximum amplitude of the velocity field
is $\delta u \approx 0.2 c_s$, where $c_s$ is the sound velocity. 
Dark features represent anticyclonic vorticity perturbations
and white features represent cyclonic vorticity perturbations. 
}} \\  

During the first few orbits all the cyclonic vorticity perturbations
were sheared by the flow and dissipated, while the anticyclonic 
vorticity perturbations became organized into vortices which later 
merged together
to form larger vortices. After 15 orbits,
only a few large elongated anticyclonic
vortices were left in the flow (Figure 9).   


{\small{\it Figure 9: A grayscale of the vorticity at 
$t \approx 15$ orbits. Anticyclonic vortices have developed and
merged 
together to form large elliptical vortices, while the cyclonic
vorticity perturbations have dissipated. }} \\ 

However, all the vortices were eventually sheared and dissipated due to
the viscosity and the shear in the flow. The perturbations did not
develop into self-sustained turbulence.  
In this model, the vortices had typical exponential decay timescales
of the order of 50 orbits 
and they can therefore survive in the
flow for a few hundreds of orbits.  
Detailed calculations of the stability and lifetime
of vortices in Keplerian disks (Godon \& Livio, 1999a),
indicate that the exponential decay time of the amplitude $A$ of the vortices
is inversely proportional to the alpha viscosity parameter.
Namely, $A \propto e^{-t/\tau}$, where $\tau \propto \alpha^{-1}$.  
Therefore, we find that in the 
range $\alpha=10^{-4} - 10^{-3}$, the decay time of the vortices 
is determined by the viscous dissipation that is introduced explicitly 
into the equations (i.e. the Reynolds stress tensor). \\ 

The decreasing kinetic energy of the fluctuations
is another indication for the stability of the flow, 
and it was observed in all the models. 
In figure 10 we show the decrease of the kinetic energy 
as a function of time for one of the models (model 4). \\  

It is also interesting to examine the kinetic energy spectrum of
the fluctuations. The energy spectrum is the same in all the models
and it is shown in Figure 11 for model 4. 
The slope of the spectrum in the short wavelengths 
(large $k$) is very
close to -3, due to the viscous dissipation of the modes. 
The rest of the spectrum (for the long wavelengths, i.e. small
$k$) is fairly flat, 
in agreement with simulations of two-dimensional
compressible turbulence (e.g. Farge \& Sadourny 1989) 
and with the turbulence obtained  
in the inner region of the disk (see \S 4.4).
The fluctuations are characterized by an inverse 
cascade of energy typical  
of two-dimensional flows, 
and as a consequence, the energy accumulates in the 
lowest (m=1) mode. The spectrum shows clearly the dominance 
of the m=1 (wavenumber $k=1$, i.e. $Log(k)=0$) mode. 
Similar energy spectra
were obtained for all the other models in this class.


{\small{\it Figure 10: The kinetic energy of the fluctuations is shown
as a function of time. 
The energy
is given on a logarithmic scale in units of the background Keplerian
energy of the flow. The energy decreases steadily with time, with
small fluctuations appearing towards the end of the run.
The same decrease of energy was observed for all the models.
The model shown here is model 4.  
}}  \\

 
{\small{\it Figure 11: The spectrum of the kinetic energy of the
fluctuations has been integrated in space and averaged over the 
last orbits. The high
frequencies are damped by the viscosity 
with a slope approaching -3. 
The rest of the spectrum is fairly flat in agreement  
with previous results for compressible 2D turbulent rotating flows.  
The low order mode m=1 has become dominant 
due to the inverse cascade of energy,
characteristic of two-dimensional turbulence. 
The same spectrum of energy was observed for all the models.
The model shown here is model 4.  
}}  \\

\subsection{Finite perturbations and tidal effects} 

As we explained in the Introduction, it is expected that the
epicyclic term will always act as a sink in the energy equation
(see \S 4.3),
thus causing the stresses to decay. It is important however to
explore whether other effects could serve as sources of energy
for the fluctuations. In principle at least, such sources
could offset the effect of the epicyclic term, with a resultant
growth instead of decay. One such potential source could be tidally 
induced spiral density waves (e.g. Frank, King \& Raine 1999). 
We have therefore run a model in 
which a finite perturbation was applied to a disk containing
a tidally induced wave. We found that the shock waves 
did not act as a source, rather, they removed energy from
the fluctuations, causing a yet faster stabilization.  

\subsection{Artificial change of the epicyclic term} 

In order to further test the stabilizing effect of the 
epicyclic term (Balbus et al. 1996, eq.2.6b), 
we introduced the following artificial
change of the epicyclic term in the angular momentum equations. 
In the angular momentum equation (eq.3), the second term on the left
hand side can be written as:
\begin{equation}
\frac{1}{r} \frac{\partial}{\partial r} r 
\left( \frac{UW}{\rho} \right) =  
\frac{1}{r} \frac{\partial}{\partial r} r
( \rho v_r v_{\phi} ) . 
\end{equation}  
We introduce a departure from the circular flow using
the relation $ u_{\phi} = v_{\phi} -r \Omega $ , 
and obtain
\begin{equation}  
\frac{1}{r} \frac{\partial}{\partial r} r 
\left( \frac{UW}{\rho} \right) = 
\frac{1}{r} \frac{\partial}{\partial r} r 
( \rho v_r u_{\phi} ) + r \Omega \frac{\partial}{\partial r} 
(\rho v_r ) + \frac{\partial}{\partial r} (r^2 \Omega) \frac{\rho v_r}
{r} .
\end{equation}  
The last term on the right hand side is the epicyclic 'sink' term described
in Balbus et al. (1996). Since in our simulations the disk
is very nearly Keplerian, we can substitute $\Omega = \Omega_K$ in
the epicyclic
term and obtain for the epicyclic term $\frac{1}{2} \rho v_r \Omega$.   
In order to modify this term in the angular momentum equation
(eq.3) 
we added on the right hand
side a term of the form $a \rho v_r r$, where $a$ is a 
positive constant. The value $a=0.5$ corresponds to a 
precise cancellation of the (sink) epicyclic term. We found
that while models with $a \le 0.5$ were stable, even a 
slight increase in $a$ above $0.5$ (e.g. $a=0.501$) resulted
in a violent instability already in one-dimensional models (used
to construct the two-dimensional models). The result was
similar to the one we obtained for Rayleigh unstable rotation
laws (specific angular momentum not increasing outwards). \\ 

The modification of the epicyclic term has no astrophysical meaning
and it was introduced only to test the hypothesis
that this term is in fact the sink term
that stabilizes the flow.

\subsection{Reflective inner boundary} 

We also performed additional tests of a phenomenon already found
by Godon (1998). In the latter study, 
a transition to turbulence
was obtained in the inner region of the disk, due to the 
development of a high-order mode of the Papaloizou-Pringle 
instability (Papaloizou \& Pringle 1984) in the non-linear regime. 
The Papaloizou-Pringle instability develops in the inner region
of the disk when the inner boundary is reflective (see also
Gat \& Livio 1992). In our current simulations, a 
transition to turbulence was observed when the 
viscosity in the disk was taken to be very low ($\alpha \approx 0.001$), 
and the order of the unstable mode was very high ($m\approx 20$).
The turbulent inner disk is shown in Figure 12. In this particular
model the resolution is $256 \times 64$.  \\  

In realistic astrophysical configurations, a 
reflective inner boundary could be obtained (i) either at the 
inner edge of the disk where a sharp drop in the density in the 
boundary layer can reflect waves (e.g. Papaloizou \& Stanley 1986), or (ii) 
at the outer layers of a white
dwarf rotating near break up in Cataclysmic Variable systems
(e.g. Narayan \& Popham 1989),
where evidence of a boundary layer is missing. 


{\small{\it Figure 12:
A grayscale of the density shows the development of turbulence
in the inner region of a thin Keplerian disk. This non-linear 
instability is global, due to the inner reflective boundary. 
}}  \\

\section{Discussion} 

Using a high-resolution, high-accuracy spectral method we have
performed two-dimensional simulations of {\it an entire} accretion disk.
The Keplerian (compressible) flow was
subjected to a variety of 
finite-amplitude perturbations. Our simulations show
that such purely hydrodynamic disturbances do not evolve into a 
self-sustained turbulence. We have shown that, for small perturbations, 
sheared
disturbances develop, but they are unable to tap energy from the flow
and they subsequently decay. Furthermore, we have shown that 
the presence of a tidal field does not enhance the turbulence but
rather suppresses it.  
For {\it large size and amplitude} 
vorticity perturbations, coherent anticyclonic
vortices form and merge together. However, here too the disturbance 
does not tap energy from the flow: the vorticity eventually decays and 
stretches due to the viscosity and shear (respectively). 
Similar results were
obtained (Bracco et al. 1998, 1999) using the incompressible
shallow water equations for the disk. In their models, 
Bracco et al. were limited to a polytropic index
$n=1$, a constant density profile $\rho$, and most importantly
the value of $H/r$ was not specified (Provenzale 1998, private
communications). In the present work we provide a more realistic
approach and confirm the results of Bracco et al. (1998, 1999).
The relatively long-lived vortices may have some important consequences
for the 
formation of giant planets in protoplanetary disks (Adams 
\& Watkins 1995; Tanga et al. 1996; Godon \& Livio 1999a) 
by allowing dust particles
to rapidly concentrate in their core. However, they do not 
lead to turbulence and angular momentum transport.  
More details on the stability and lifetime of vortices can
be found in Godon \& Livio (1999a).  \\  

We have also confirmed the fact that the epicyclic term in the energy
equation, which acts as a sink, is the main cause for the 
stability of Keplerian disks. Our calculations have shown that the transfer
of energy to the largest scale results in a (transient) dominance 
of the m=1 mode. This could have important consequences for 
short period oscillations of disks (Godon \& Livio 1999b). 
Finally, we have shown that when an inner reflective boundary
is assumed, the inner disk is globally non-linearly unstable
to non-axisymmetric modes which can evolve towards turbulence
(see also Godon 1998; and arguments by Brandenburg et al. 1995). 
While this is an interesting theoretical result, we do not
think that this instability plays a role in global angular
momentum transport in disks, since the turbulence is confined 
to the very inner disk. In addition, the relevance of an inner reflective
boundary condition has been only partially justified in some systems.  
\\ 
To conclude, while a definitive answer will have to await for full-scale
three-dimensional calculations, our results strengthen the impression
that purely hydrodynamical instabilities are probably not the source of
angular momentum transport in accretion disks. MHD turbulence on the other
hand not only transports angular momentum adequately, but also readily
lends itself to the $\alpha$-disk formalism (Balbus \& Papaloizou 1999). 
 
\section*{Acknowledgments} 

ML acknowledges support from NASA Grant NAG5-6857. 
This work was supported in part by the Director's Discretionary
Research Fund at Space Telescope Science Institute. 


\section*{References}

\noindent       
Adam, F. C., \& Watkins, R. 1995, ApJ, 451, 314 
\\ \\          
Balbus, S. A., \& Hawley, J. F. 1991, ApJ, 376, 214
\\ \\          
Balbus, S. A., \& Hawley, J. F. 1998, Rev. Mod. Phys., 70, 1  
\\ \\          
Balbus, S. A., Hawley, J. F., \& Stone, J. M. 1996, ApJ, 467, 76 
\\ \\          
Balbus, S. A., \& Papaloizou, J. C. B., ApJ, submitted 
\\ \\          
Bayly, B. J., Orszag, S. A., \& Herbert, T. 1988, Ann.Rev.Fluid Mech., 20, 359 
\\ \\          
Blaes, O. M., \& Balbus, S. A. 1994, ApJ, 421, 163 
\\ \\     
Bracco, A., Provenzale, A., Spiegel, E., Yecko, P., 1998, in
Abramowicz A. (Ed.), Proceedings of the conference on Quasars and 
Accretion Disks. Cambridge Univ. Press 
\\ \\
Bracco, A., Chavanis, P. H., Provenzale, A., Spiegel, E. A., 1999,
Physics of Fluids, in press 
\\ \\       
Brandenburg, A., Nordlund, A., Stein, R. F., \& Torkelsson, U. 1995,
ApJ, 446, 741 
\\ \\          
Canuto, C., Hussaini, M. Y., Quarteroni, A. \& Zang, T. A. 1988, 
Spectral Methods in Fluid Dynamics (New York: Springer Verlag) 
\\ \\          
Chandrasekhar, S. 1960, Proc. Nat. Acad. Sci., 46, 253 
\\ \\          
Cho, J. Y. K., \& Polvani, L. M. 1996a, Phys. Fluids, 8 (6), 1531 
\\ \\          
Cho, J. Y. K., \& Polvani, L. M. 1996b, Science, 273, 335  
\\ \\          
Coles, D. 1965, J. Fluid Mech., 21, 385 
\\ \\          
Cushman-Roisin, B. 1994, Introduction to Geophysical Fluid Dynamics,
(Prentice-Hall, New Jersey)
\\ \\          
Daviaud, F., Hegseth, J., \& Berg\'e, P. 1992, Phys.Rev.Let., 69 (17), 2511 
\\ \\          
Dauchot, O., \& Daviaud, F. 1994, Europhys. Lett., 28, 225 
\\ \\          
Dauchot, O., \& Daviaud, F. 1995, Phys.Fluids, 7 (5), 901  
\\ \\          
Don, W. S., \& Gottlieb, D. 1990, Comput. Meth. Appl. Mech. Engrg., 80, 39 
\\ \\          
Dubrulle, B. 1991, in P.Sulem \& J.-D.Fournier (eds.), Non linear
dynamics of structures, (Springer Verlag), 252 
\\ \\          
Dubrulle, B. 1992, A \& A, 266, 592 
\\ \\          
Dubrulle, B. 1993, ICARUS, 106, 59  
\\ \\          
Dubrulle, B., \& Knobloch, E. 1992, A \& A, 256, 673 
\\ \\          
Dubrulle, B., \& Nazarenko, S. 1994, Europhys. Lett., 27, 129  
\\ \\          
Dubrulle, B., \& Valdettaro, L. 1992, A \& A, 263, 387   
\\ \\          
Farge, M., \& Sadourny, R. 1989, J. Fluid Mech., 206, 433 
\\ \\          
Fj{\o}rtoft, R. 1950, Geofys. Publ., 17, 52
\\ \\          
Frank, J., King, A. R., \& Raine, D. J. 1999, in Accretion Power in 
Astrophysics (Cambridge: Cambridge University Press), in press 
\\ \\          
Gat, O., \& Livio, M. 1992, ApJ, 396, 542 
\\ \\          
Godon, P. 1997, ApJ, 480, 329 
\\ \\          
Godon, P. 1998, ApJ, 502, 382 
\\ \\          
Godon, P., \& Livio M., 1999a, submitted to ApJ   
\\ \\          
Godon, P., \& Livio M., 1999b, in preparation  
\\ \\          
Gottlieb, D., \& Orszag, S.A. 1977, Numerical Analysis of Spectral
Methods: Theory and Applications (NSF-CBMS Monograph 26;
Philadelphia: SIAM) 
\\ \\ 
Hamilton, J. M., \& Abernathy, F. H. 1994, J. Fluid Mech., 264, 185 
\\ \\       
Hawley, J. F., Gammie, C. F., Balbus, S. A. 1995, ApJ, 440, 742 
\\ \\       
Hegseth, J. J. 1996, Phys. Rev. E, 54 (5), 4915 
\\ \\       
Narayan, R., \& Popham, R. 1989, ApJ, 346, L25  
\\ \\       
Orszag, S. A., \& Kells, L. C. 1980, J.Fluid Mech., 96, 159 
\\ \\       
Papaloizou, J. C. B., \& Pringle, J. E., 1984, MNRAS, 208, 721   
\\ \\       
Papaloizou, J. C. B., \& Stanley, G. Q. G., 1986, MNRAS, 220, 593   
\\ \\       
Pedlosky, J. 1987, Geophysical Fluid Dynamics, 2nd ed.
(Springer Verlag, New York)
\\ \\ 
Reg\"os, E. 1997, MNRAS, 286, 97 
\\ \\       
Savonije,  G. J., Papaloizou, J. C. B., \& Lin, D. N. C., 1994,
MNRAS, 268, 13 
Shakura, N.I., \& Sunyaev, R. A. 1973, A \& A, 24, 337. 
\\ \\       
She, Z.S., Jackson, E., \& Orszag, S.A. 1991, Proceedings of the Royal
Society of London, Series A: Mathematical and Physical Sciences,
vol. 434 (1890), 101. 
\\ \\       
Tanga, P., Babiano, A., Dubrulle, B., \& Provenzale, A., 1996,
ICARUS, 121, 158 
\\ \\       
Tomasini, M., Dolez, N., \& L\'eorat, J. 1996, J. Fluid Mech., 306, 59  
\\ \\       
Velikhov, E. P. 1959, Soviet Phys. - JETP, 36, 1398 
\\ \\       
Vishniac, E. T., \& Brandenburg, A. 1997, ApJ, 475, 263 
\\ \\       
Voigt, R.G., Gottlieb, D., \& Hussaini, M.Y. 1984, Spectral Methods for
Partial Differential Equations (Philadelphia: SIAM-CBMS) 
\\ \\      
Yang, Z. 1987, PhD thesis, Division of Applied Sciences, Harvard
University
\\ \\      
Zahn, J. P. 1990, in Bertout, C., Collin-Souffrin, S., Lasota, J. P., 
\& Tran Than, Van J., (eds.), Structure and emission properties of 
accretion disks, Edition Fronti\`eres

\end{document}